%% file: B-L_unitarity.tex
\documentclass[12pt]{article}
\usepackage{graphicx}
\usepackage{epstopdf}
\usepackage{wrapfig}
\usepackage{subfig}
\usepackage{caption}
\usepackage{mathrsfs}
\usepackage{amssymb}
\usepackage{amsmath}
\usepackage{verbatim}
\advance\textheight by \topskip
\begin{document}


\newcommand{\HPA}[1]{{\it Helv.\ Phys.\ Acta.\ }{\bf #1}}
\newcommand{\AP}[1]{{\it Ann.\ Phys.\ }{\bf #1}}
\newcommand{\be}{\begin{equation}}
\newcommand{\ee}{\end{equation}}
\newcommand{\br}{\begin{eqnarray}}
\newcommand{\er}{\end{eqnarray}}
\newcommand{\ba}{\begin{array}}
\newcommand{\ea}{\end{array}}
\newcommand{\bi}{\begin{itemize}}
\newcommand{\ei}{\end{itemize}}
\newcommand{\bn}{\begin{enumerate}}
\newcommand{\en}{\end{enumerate}}
\newcommand{\bc}{\begin{center}}
\newcommand{\ec}{\end{center}}
\newcommand{\ul}{\underline}
\newcommand{\ol}{\overline}
\def\l{\left\langle}
\def\r{\right\rangle}
\def\as{\alpha_{s}}
\def\ycut{y_{\mbox{\tiny cut}}}
\def\yij{y_{ij}}
\def\epem{\ifmmode{e^+ e^-} \else{$e^+ e^-$} \fi}
\newcommand{\eeww}{$e^+e^-\rightarrow W^+ W^-$}
\newcommand{\qqQQ}{$q_1\bar q_2 Q_3\bar Q_4$}
\newcommand{\eeqqQQ}{$e^+e^-\rightarrow q_1\bar q_2 Q_3\bar Q_4$}
\newcommand{\eewwqqqq}{$e^+e^-\rightarrow W^+ W^-\ar q\bar q Q\bar Q$}
\newcommand{\eeqqgg}{$e^+e^-\rightarrow q\bar q gg$}
\newcommand{\eeqloop}{$e^+e^-\rightarrow q\bar q gg$ via loop of quarks}
\newcommand{\eeqqqq}{$e^+e^-\rightarrow q\bar q Q\bar Q$}
\newcommand{\eewwjjjj}{$e^+e^-\rightarrow W^+ W^-\rightarrow 4~{\rm{jet}}$}
\newcommand{\eeqqggjjjj}{$e^+e^-\rightarrow q\bar 
q gg\rightarrow 4~{\rm{jet}}$}
\newcommand{\eeqloopjjjj}{$e^+e^-\rightarrow q\bar 
q gg\rightarrow 4~{\rm{jet}}$ via loop of quarks}
\newcommand{\eeqqqqjjjj}{$e^+e^-\rightarrow q\bar q Q\bar Q\rightarrow
4~{\rm{jet}}$}
\newcommand{\eejjjj}{$e^+e^-\rightarrow 4~{\rm{jet}}$}
\newcommand{\jjjj}{$4~{\rm{jet}}$}
\newcommand{\qqbar}{$q\bar q$}
\newcommand{\ww}{$W^+W^-$}
\newcommand{\ar}{\rightarrow}
\newcommand{\sm}{${\cal {SM}}$}
\newcommand{\Dir}{\kern -6.4pt\Big{/}}
\newcommand{\Dirin}{\kern -10.4pt\Big{/}\kern 4.4pt}
\newcommand{\DDir}{\kern -8.0pt\Big{/}}
\newcommand{\DGir}{\kern -6.0pt\Big{/}}
\newcommand{\wwqqqq}{$W^+ W^-\ar q\bar q Q\bar Q$}
\newcommand{\qqgg}{$q\bar q gg$}
\newcommand{\qloop}{$q\bar q gg$ via loop of quarks}
\newcommand{\qqqq}{$q\bar q Q\bar Q$}

\def\st{\sigma_{\mbox{\scriptsize t}}}
\def\Ord{\buildrel{\scriptscriptstyle <}\over{\scriptscriptstyle\sim}}
\def\OOrd{\buildrel{\scriptscriptstyle >}\over{\scriptscriptstyle\sim}}
\def\jhep #1 #2 #3 {{JHEP} {\bf#1} (#2) #3}
\def\plb #1 #2 #3 {{Phys.~Lett.} {\bf B#1} (#2) #3}
\def\npb #1 #2 #3 {{Nucl.~Phys.} {\bf B#1} (#2) #3}
\def\epjc #1 #2 #3 {{Eur.~Phys.~J.} {\bf C#1} (#2) #3}
\def\zpc #1 #2 #3 {{Z.~Phys.} {\bf C#1} (#2) #3}
\def\jpg #1 #2 #3 {{J.~Phys.} {\bf G#1} (#2) #3}
\def\prd #1 #2 #3 {{Phys.~Rev.} {\bf D#1} (#2) #3}
\def\prep #1 #2 #3 {{Phys.~Rep.} {\bf#1} (#2) #3}
\def\prl #1 #2 #3 {{Phys.~Rev.~Lett.} {\bf#1} (#2) #3}
\def\mpl #1 #2 #3 {{Mod.~Phys.~Lett.} {\bf#1} (#2) #3}
\def\rmp #1 #2 #3 {{Rev. Mod. Phys.} {\bf#1} (#2) #3}
\def\cpc #1 #2 #3 {{Comp. Phys. Commun.} {\bf#1} (#2) #3}
\def\sjnp #1 #2 #3 {{Sov. J. Nucl. Phys.} {\bf#1} (#2) #3}
\def\xx #1 #2 #3 {{\bf#1}, (#2) #3}
\def\hepph #1 {{\tt hep-ph/#1}}
\def\preprint{{preprint}}

\def\beq{\begin{equation}}
\def\beeq{\begin{eqnarray}}
\def\eeq{\end{equation}}
\def\eeeq{\end{eqnarray}}
\def\a0{\bar\alpha_0}
\def\thrust{\mbox{T}}
\def\Thrust{\mathrm{\tiny T}}
\def\ae{\alpha_{\mbox{\scriptsize eff}}}
\def\ap{\bar\alpha_p}
\def\as{\alpha_{\mathrm{S}}}
\def\aem{\alpha_{\mathrm{EM}}}
\def\b0{\beta_0}
\def\cN{{\cal N}}
\def\cd{\chi^2/\mbox{d.o.f.}}
\def\Ecm{E_{\mbox{\scriptsize cm}}}
\def\ee{e^+e^-}
\def\enap{\mbox{e}}
\def\eps{\epsilon}
\def\ex{{\mbox{\scriptsize exp}}}
\def\GeV{\mbox{\rm{GeV}}}
\def\half{{\textstyle {1\over2}}}
\def\jet{{\mbox{\scriptsize jet}}}
\def\kij{k^2_{\bot ij}}
\def\kp{k_\perp}
\def\kps{k_\perp^2}
\def\kt{k_\bot}
\def\lms{\Lambda^{(n_{\rm f}=4)}_{\overline{\mathrm{MS}}}}
\def\mI{\mu_{\mathrm{I}}}
\def\mR{\mu_{\mathrm{R}}}
\def\MSbar{\overline{\mathrm{MS}}}
\def\mx{{\mbox{\scriptsize max}}}
\def\NP{{\mathrm{NP}}}
\def\pd{\partial}
\def\pt{{\mbox{\scriptsize pert}}}
\def\pw{{\mbox{\scriptsize pow}}}
\def\so{{\mbox{\scriptsize soft}}}
\def\st{\sigma_{\mbox{\scriptsize tot}}}
\def\ycut{y_{\mathrm{cut}}}
\def\slashchar#1{\setbox0=\hbox{$#1$}           
     \dimen0=\wd0                                 
     \setbox1=\hbox{/} \dimen1=\wd1               
     \ifdim\dimen0>\dimen1                        
        \rlap{\hbox to \dimen0{\hfil/\hfil}}      
        #1                                        
     \else                                        
        \rlap{\hbox to \dimen1{\hfil$#1$\hfil}}   
        /                                         
     \fi}                                         %
\def\etmiss{\slashchar{E}^T}
\def\Meff{M_{\rm eff}}
\def\Ord{\lsim}
\def\OOrd{\gsim}
\def\tq{\tilde q}
\def\tchi{\tilde\chi}
\def\lsp{\tilde\chi_1^0}

\def\gam{\gamma}
\def\ph{\gamma}
\def\be{\begin{equation}}
\def\ee{\end{equation}}
\def\bea{\begin{eqnarray}}
\def\eea{\end{eqnarray}}
\def\lsim{\:\raisebox{-0.5ex}{$\stackrel{\textstyle<}{\sim}$}\:}
\def\gsim{\:\raisebox{-0.5ex}{$\stackrel{\textstyle>}{\sim}$}\:}

\def\ino{\mathaccent"7E} \def\gluino{\ino{g}} \def\mgluino{m_{\gluino}}
\def\sqk{\ino{q}} \def\sup{\ino{u}} \def\sdn{\ino{d}}
\def\chargino{\ino{\omega}} \def\neutralino{\ino{\chi}}
\def\cab{\ensuremath{C_{\alpha\beta}}} \def\proj{\ensuremath{\mathcal P}}
\def\dab{\delta_{\alpha\beta}}
\def\zz{s-M_Z^2+iM_Z\Gamma_Z} \def\zw{s-M_W^2+iM_W\Gamma_W}
\def\prop{\ensuremath{\mathcal G}} \def\ckm{\ensuremath{V_{\rm CKM}^2}}
\def\aem{\alpha_{\rm EM}} \def\stw{s_{2W}} \def\sttw{s_{2W}^2}
\def\nc{N_C} \def\cf{C_F} \def\ca{C_A}
\def\qcd{\textsc{Qcd}} \def\susy{supersymmetric} \def\mssm{\textsc{Mssm}}
\def\slash{/\kern -5pt} \def\stick{\rule[-0.2cm]{0cm}{0.6cm}}
\def\h{\hspace*{-0.3cm}}

\def\ims #1 {\ensuremath{M^2_{[#1]}}}
\def\tw{\tilde \chi^\pm}
\def\tz{\tilde \chi^0}
\def\tf{\tilde f}
\def\tl{\tilde l}
\def\ppb{p\bar{p}}
\def\gl{\tilde{g}}
\def\sq{\tilde{q}}
\def\sqb{{\tilde{q}}^*}
\def\qb{\bar{q}}
\def\sqL{\tilde{q}_{_L}}
\def\sqR{\tilde{q}_{_R}}
\def\ms{m_{\tilde q}}
\def\mg{m_{\tilde g}}
\def\Gs{\Gamma_{\tilde q}}
\def\Gg{\Gamma_{\tilde g}}
\def\md{m_{-}}
\def\eps{\varepsilon}
\def\Ce{C_\eps}
\def\dnq{\frac{d^nq}{(2\pi)^n}}
\def\DR{$\overline{DR}$\,\,}
\def\MS{$\overline{MS}$\,\,}
\def\DRm{\overline{DR}}
\def\MSm{\overline{MS}}
\def\ghat{\hat{g}_s}
\def\shat{\hat{s}}
\def\sihat{\hat{\sigma}}
\def\Li{\text{Li}_2}
\def\bs{\beta_{\sq}}
\def\xs{x_{\sq}}
\def\xsa{x_{1\sq}}
\def\xsb{x_{2\sq}}
\def\bg{\beta_{\gl}}
\def\xg{x_{\gl}}
\def\xga{x_{1\gl}}
\def\xgb{x_{2\gl}}
\def\lsp{\tilde{\chi}_1^0}

\def\gluino{\mathaccent"7E g}
\def\mgluino{m_{\gluino}}
\def\squark{\mathaccent"7E q}
\def\msquark{m_{\mathaccent"7E q}}
\def\M{ \overline{|\mathcal{M}|^2} }
\def\utm{ut-M_a^2M_b^2}
\def\MiLR{M_{i_{L,R}}}
\def\MiRL{M_{i_{R,L}}}
\def\MjLR{M_{j_{L,R}}}
\def\MjRL{M_{j_{R,L}}}
\def\tiLR{t_{i_{L,R}}}
\def\tiRL{t_{i_{R,L}}}
\def\tjLR{t_{j_{L,R}}}
\def\tjRL{t_{j_{R,L}}}
\def\tg{t_{\gluino}}
\def\uiLR{u_{i_{L,R}}}
\def\uiRL{u_{i_{R,L}}}
\def\ujLR{u_{j_{L,R}}}
\def\ujRL{u_{j_{R,L}}}
\def\ug{u_{\gluino}}
\def\utot{u \leftrightarrow t}
\def\ar{\to}
\def\sqk{\mathaccent"7E q}
\def\sup{\mathaccent"7E u}
\def\sdn{\mathaccent"7E d}
\def\chargino{\mathaccent"7E \chi}
\def\neutralino{\mathaccent"7E \chi}
\def\slepton{\mathaccent"7E l}
\def\M{ \overline{|\mathcal{M}|^2} }
\def\cab{\ensuremath{C_{\alpha\beta}}}
\def\ckm{\ensuremath{V_{\rm CKM}^2}}
\def\zz{s-M_Z^2+iM_Z\Gamma_Z}
\def\zw{s-M_W^2+iM_W\Gamma_W}
\def\s22w{s_{2W}^2}

\newcommand{\cpmtwo}    {\mbox{$ {\chi}^{\pm}_{2}                    $}}
\newcommand{\cpmone}    {\mbox{$ {\chi}^{\pm}_{1}                    $}}

\begin{flushright}
{SHEP-10-26}\\
\today
\end{flushright}
\vskip0.1cm\noindent
\begin{center}
{{\Large {\bf Constraining the $g'_1$ coupling \\[0.25cm] in the
      minimal $B-L$ Model}}
\\[1.0cm]
{\large Lorenzo Basso, Stefano Moretti and Giovanni Marco Pruna}\\[0.30
  cm]
{\it  School of Physics and Astronomy, University of Southampton,}\\
{\it  Highfield, Southampton SO17 1BJ, UK.}
}
\\[1.25cm]
\end{center}

\begin{abstract}
{\small
\noindent
We have combined perturbative unitarity and renormalisation group
equation arguments in order to find a dynamical way to constrain the
$g'_1$ coupling of the minimal $B-L$ extension of the Standard
Model. We have analysed the role of the $g'_1$ coupling evolution in
the perturbative stability of the two-to-two body scattering
amplitudes of the vector boson and scalar sectors of the model and
we have shown that perturbative unitarity imposes an upper bound
on the $B-L$ gauge coupling. We have made a comparison between this
criterion and the triviality arguments, showing that our procedure
substantially refines the triviality bounds.
}

\end{abstract}


\section{Introduction}
\label{Sec:Intro}
\input{sect_1.tex}




\section{Constraining the $g'_1$ of the minimal $B-L$ model}
\label{Sec:bounds}
\input{sect_3.tex}


\section{Results}
\label{Sec:Results}
\input{sect_4.tex}


\section{Conclusions}
\label{Sec:Conclusions}
\input{sect_5.tex}


\section*{Acknowledgements} 
\input{acknowledgements.tex}




\input{bibl.tex}
\end{document}

%% file: sect_1.tex
Nowadays the phenomenological importance of Beyond the Standard Model
(BSM) physics at the TeV scale is recognised by the global
experimental effort at the Large Hadron Collider (LHC).

The minimal $B-L$ (baryon 
minus lepton number) extension of the Standard Model (SM) \cite{B-L} is
considered as one of the 
candidates in the description of a promising and simple
BSM scenario, containing a significant set of particles and interactions
whose existence could be proven both at the LHC (see \cite{B-L:Khal,B-L:LHC}) and future Linear Colliders (LCs) \cite{bbmp}.

This model is based on the gauging of the $B-L$ symmetry: one obtains
said extension of the SM by augmenting the gauge groups with an additional
$U(1)_{B-L}$: $SU(3)_C\times SU(2)_L\times U(1)_Y\times U(1)_{B-L}$. 

For consistency, three generations of heavy neutrinos in the
fermion sector to cure anomalies and a complex singlet scalar field
 must be included, the latter giving rise to an extra Higgs boson
after the spontaneous symmetry breaking of the new gauge group.

In all generality, the two $U(1)$ gauge groups will mix together, giving raise to a set of two new gauge couplings, $g'_1$ and $\widetilde{g}$. While the former appears in the covariant derivative as purely related to the $B-L$ charge, the latter (coupling the new B-L gauge field to the hypercharge) controls the mixing between the two neutral massive gauge bosons at the tree level.

However, from LEP analysis \cite{Carena:2004xs} it is known that once a Tera-scale $Z'$ is considered (as it is the case of this paper), the small mixing observed between $Z$ and $Z'$ could be realised only by means of a small $\widetilde{g}$ coupling. Hence, as a reasonable approximation, we decided to switch off the $\widetilde{g}$ coupling, concentrating on the ``pure'' $B-L$ model only.
This choice allows to perform an analytic analysis, otherwise precluded when full $Z-Z'$ mixing is taken into account.

The parameter space arising from the $B-L$ extension is bounded by
both experimental (mainly precision tests at LEP,
see \cite{Cacciapaglia:2006pk}) and theoretical arguments. For the
latter, a recent set of works (see \cite{Basso:2010jt}
and \cite{Basso:2010jm}) has been devoted to constrain the scalar
sector
and the $g'_1$ coupling, that is, the only new gauge coupling of
the minimal $B-L$ model at the EW scale (since, as intimated, no
mixing is allowed between the SM $Z$ and $B-L$ $Z'$ boson at tree-level
at such a scale).

The purpose of this paper is to show that
the renormalisation group equation (RGE) based techniques as well as the perturbative
unitarity criterion can be combined to give a dynamical way to
constrain the $g'_1$ coupling.

To this end, we propose a detailed study of the vector
boson and Higgs sectors of the model with a view to extract the most stringent
bound on the (evolving) $g'_1$ coupling.
We will make a comparison between this method and triviality
arguments, showing that calling for perturbative unitarity
stability allows us to obtain a stronger constraint on $g'_1$ with respect to
traditional triviality assumptions. 

This work is organised as follows: in section \ref{Sec:bounds} we
describe the theoretical methods adopted to constrain the
$g'_1$ coupling, in section \ref{Sec:Results} we present
our numerical results while in the last section we give our conclusions.

%% file: sect_3.tex
The model under study is the so-called ``pure'' or ``minimal''
$B-L$ model (see \cite{B-L:LHC} for conventions and references) 
since it has vanishing mixing between the two $U(1)_{Y}$ 
and $U(1)_{B-L}$ gauge groups.
In the rest of this paper we refer to this model simply as the ``$B-L$
model''.  In this scenario the classical gauge invariant Lagrangian,
obeying the $SU(3)_C\times SU(2)_L\times U(1)_Y\times U(1)_{B-L}$
gauge symmetry, can be decomposed as:
\begin{equation}\label{L}
\mathscr{L}=\mathscr{L}_{YM} + \mathscr{L}_s + \mathscr{L}_f
+ \mathscr{L}_Y \, ,
\end{equation}
where $\mathscr{L}_{YM}$, $\mathscr{L}_s$, $\mathscr{L}_f$ and
$\mathscr{L}_Y$ are the Yang-Mills, scalar, fermionic and Yukawa
sectors, respectively.
Since it has been proven that perturbative unitarity violation at high 
energy occurs only in 
vector and Higgs bosons elastic scatterings, our 
interest is focused on the corresponding sectors.

Following the Becchi-Rouet-Stora (BRS) invariance (see \cite{BecchiRouetStora}),  the amplitude for
emission or absorption of a ``scalarly'' polarised gauge
boson becomes equal to the amplitude 
for emission or absorption of the related would-be-Goldstone boson,
and, in the high energy limit ($s \gg m^2_{W^{\pm},Z,Z'}$), the
amplitude involving the (physical) longitudinal polarisation (the dominant one) of
gauge bosons approaches the (unphysical) scalar one, the so-called
Equivalence Theorem (ET), see \cite{ChanowitzGaillard85}. Therefore, the analysis of the
perturbative unitarity of two-to-two
particle scatterings in the gauge sector can be performed, in the high
energy limit, by exploiting the Goldstone sector. (Further details of
this formalism can be found in \cite{Basso:2010jt}.)


Moreover, since we want to focus on $g'_1$ limits,
we assume that the two Higgs bosons of the model have masses such that no 
significant contribution to the spherical partial wave amplitude will come 
from the scalar four-point and three-point functions, 
according to \cite{Basso:2010jt}). Such upper value is usually referred as the Lee-Quigg-Tacker (LQT) 
limit \cite{LeeQuiggThacker} on the Higgs boson mass, evaluated in the ET framework. 
Taking Higgs boson masses smaller than the LQT limit is therefore a way to exclude any other source of 
unitarity violation different from the largeness of the $g'_1$ gauge coupling.

In the search for the $g'_1$ upper limits, we will
assume that we can neglect the other gauge couplings in the covariant
derivative:
\begin{equation}\label{cov_der}
D_{\mu}\simeq \partial _{\mu} + i g_1'Y^{B-L}Z'_{\mu}\,.
\end{equation}
In order to have a consistently gauge invariant theory, in this
particular model we must
choose $Y^{B-L}_{H}=0$ and $Y^{B-L}_{\chi}=2$\footnote{In other
versions of the $B-L$ model this quantum number could change: for example, 
in \cite{Khalil:2010zz} one has $Y^{B-L}_{\chi}=1$.}, and this leads
us to a relatively small set of Feynman Rules (FRs) for the Higgs and
Goldstone sector of the theory.

The scalar Lagrangian and its FRs have been thoroughly
studied in \cite{Basso:2010jt}, where it was shown that the inclusion of
$g'_1$ in the covariant derivative gives rise to two new FRs, i.e.
\begin{eqnarray}
Z'h_1z' &=& -iY^{B-L}_{\chi}g'_1\sin{\alpha}(p_{h_1}^\mu-p_{z'}^\mu), \\
Z'h_2z' &=& \phantom{-} i Y^{B-L}_{\chi} g'_1 \cos{\alpha}
(p_{h_2}^\mu-p_{z'}^\mu),
\end{eqnarray}
where all the momenta are considered incoming and $z'$ is the would-be-Goldstone 
boson associated with the new $Z'$ gauge field. 

Finally, it is important to recall the relation between the $Z'$ mass
and the $B-L$ Higgs singlet Vacuum Expectation Value(VEV) $x$, that is,
\begin{eqnarray}\label{vev}
M_{Z'}=Y^{B-L}_{\chi}g'_1x.
\end{eqnarray}

Now that the background is set, we focus on the techniques that
we have used to obtain the aforementioned unitarity
bounds in combination with the RGE analysis.

Firstly, it is crucial to define the evolution of the gauge couplings
via the RGEs and their boundary
conditions.
As already established in \cite{Basso:2010jm}, the RGEs of $g_1$,
$g'_1$ and $\widetilde g$ are:
\begin{eqnarray}\label{RGE_g1}
\frac{d(g_1)}{d(\log{\Lambda})} &=& \frac{1}{16\pi
^2}\left[\frac{41}{6}g_1^3 \right]\, , \nonumber \\ \label{RGE_g2} 
\frac{d(g_1')}{d(\log{\Lambda})} &=& \frac{1}{16\pi
^2}\left[ \frac{32+(Y^{B-L}_{\chi})^2}{3} g_1'^3 +
2 \frac{16}{3} g_1'^2\widetilde g+\frac{41}{6}g_1'\widetilde
g^2 \right] \, , \\ \label{RGE_g_tilde} 
\frac{d(\widetilde g)}{d(\log{\Lambda})}&=& \frac{1}{16\pi
^2}\left[\frac{41}{6}\widetilde{g}\,(\widetilde
g^2+2g_1^2) + 2 \frac{16}{3} g_1' (\widetilde{g}^2 + g_1^2)
+ \frac{32+(Y^{B-L}_{\chi})^2}{3} g_1'^2\widetilde
g \right]\, , \nonumber 
\end{eqnarray}
where $g_1(EW)\simeq 0.36$ and $\widetilde{g}(EW)=0$ (in the minimal
$B-L$ model there is no mixing at the EW scale).
This fully fixes the evolution of $g'_1$ with the scale.

In the search for the maximum $g'_1(EW)$ allowed by theoretical
constraints, the contour condition
\begin{eqnarray}\label{triviality_condition}
g'_1(\Lambda)\leq k,
\end{eqnarray}
also known as the triviality condition, is the assumption that enables to solve the system of eqs. and gives the traditional upper bound on $g'_1$ at the EW scale.

It is usually assumed either $k=1$ or $k=\sqrt{4\pi}$, calling for a
coupling that preserves the perturbative convergence of the theory.
Nevertheless, we stress again that this is an ``ad hoc''
assumption. 
Our aim, instead, is to extract the boundary condition by perturbative
unitarity arguments, showing that under certain conditions it
represents a stronger constraint on the domain of $g'_1$.

For this, we exploit the theoretical techniques that are related with
the perturbative unitarity analysis, since they can be used to provide
constraints on the theory, with a procedure that is not far from the
one firstly described in detail by \cite{LeeQuiggThacker}.

The well known result is that, by evaluating the tree-level scattering
amplitude of longitudinally polarised vector bosons, one finds that the
latter grows with the energy of the process, and in order to preserve
unitarity it is necessary to include some other (model dependent)
interactions (for example, in the SM one needs to include
the Higgs boson) and these must fulfill the unitarity criterion (again in the SM, the Higgs boson must have a mass bounded
from above by the LQT limit: $m_h\leq 700$ GeV, see \cite{LuscherWeisz}).

As already intimated, we also know that the ET allows
to compute the amplitude of any process with external longitudinal
vector bosons $V_L$ ($V = W^\pm,Z,Z' $), in the limit $m^2_V\ll s$,
by substituting each one of them with the related Goldstone boson $v
= w^\pm,z,z'$, and its general validity has been proven
in \cite{ChanowitzGaillard85}. Schematically, if we consider a
process with four longitudinal vector bosons, we have that $M(V_L V_L \rightarrow V_L V_L) = M(v v \rightarrow v v)+ O(m_V^2/s)$.

While in the search for the Higgs boson mass bound it is widely accepted
to assume small values for the gauge couplings and large Higgs boson masses, for our purpose
we reverse such argument with the same logic: we assume that
the Higgs boson masses are compatible with the unitarity limits and
we study the two-to-two scattering amplitudes of the whole scalar sector, pushing the
largeness of $g'_1$ to the perturbative limit.

This limit is a consequence of the following argument: given a
tree-level scattering amplitude between two spin-$0$ particles,
$M(s,\theta)$, where $\theta$ is the scattering (polar) angle, 
we know that the partial wave amplitude with angular
momentum $J$ is given by
\begin{eqnarray}\label{integral}
a_J = \frac{1}{32\pi} \int_{-1}^{1} d(\cos{\theta}) P_J(\cos{\theta})
M(s,\theta),
\end{eqnarray}
where $P_J$ are the Legendre polynomials, and it has been proven
(see \cite{LuscherWeisz}) that, in order to preserve unitarity, each
partial wave must be bounded by the condition
\begin{eqnarray}\label{condition}
|\textrm{Re}(a_J(s))|\leq \frac{1}{2}.
\end{eqnarray}

By direct computation, it turns out that only $J=0$ (corresponding to
the spherical partial wave contribution) leads to some bound, so we
will not discuss the higher partial waves any further.

Assuming that the Higgs boson masses do not play any role in the
perturbative unitarity violation ($m_{h_{1,2}}<700$ GeV, according to the LQT limit), we have
proven that the only divergent 
contribution to the spherical amplitude is due to the size
of the coupling $g'_1$ in the intermediate $Z'$ vector boson exchange
contributions.
Hence, the only relevant channels are: $z'z'\rightarrow h_1 h_1$,
$z'z'\rightarrow h_1 h_2$, $z'z'\rightarrow h_2 h_2$.

As an example, we evaluate the $a_0$ partial wave amplitude for
$z'z'\rightarrow h_1 h_1$ scattering in the $s\gg M_{Z'}, m_{h_{1}}$
limit.

Firstly, we know that
\begin{eqnarray}\label{mzpzph1h1}
M(s,\cos{\theta})\simeq (Y^{B-L}_{\chi} g'_1 \sin{\alpha})^2
\left(
1-\frac{4s}{s
\left(
1-\cos{\theta}
\right)+2M^2_{Z'}\cos{\theta}}
\right),
\end{eqnarray}
by the integration proposed in equation (\ref{integral}), we then extract
the $J=0$ partial wave:
\begin{eqnarray}\label{a_0z'z'h_1h_1}
a_0(z'z'\rightarrow h_1h_1)=\frac{(Y^{B-L}_{\chi}g'_1)^2}{16\pi}
\left(
1+2\log{\left( \frac{M_{Z'}^2}{s} \right)}
\right)\sin^2{\alpha}.
\end{eqnarray}

It is important to notice that the mass of the $Z'$ acts as a natural
regularisator that preserves both the amplitude and the spherical
partial wave from any collinear divergence.

Considering the three aforementioned scattering channels, their
spherical partial wave (in the
high energy limit $s\gg M_{Z'}, m_{h_{1,2}}$) is represented by the
following matrix: 
\begin{eqnarray}\label{a0_matrix}
a_0=f(g'_1,s;Y^{B-L}_{\chi},x)
\left(
\begin{tabular}{cccc}
$0$ & $\frac{1}{2}\sin^2{\alpha}$
& $-\frac{1}{\sqrt{2}} \sin{\alpha} \cos{\alpha} $ &
$\frac{1}{2}\cos^2{\alpha}$ \\
$\frac{1}{2}\sin^2{\alpha}$ & $0$ & $0$ & $0$ \\
$-\frac{1}{\sqrt{2}} \sin{\alpha} \cos{\alpha} $ & $0$ &
$0$ & $0$ \\
$\frac{1}{2}\cos^2{\alpha}$ & $0$ & $0$ & $0$ \\
\end{tabular}
\right),
\end{eqnarray}
where, according to equation (\ref{vev}),
\begin{eqnarray}\label{function}
f(g'_1,s;Y^{B-L}_{\chi},x)=\frac{(Y^{B-L}_{\chi}g'_1)^2}{16\pi}
\left(
1+2\log{\left( \frac{(Y^{B-L}_{\chi} g'_1x)^2}{s} \right)}
\right),
\end{eqnarray}
and the elements of the matrix are related to the four channels system
consisting of
$\frac{1}{\sqrt{2}}z'z'$, $\frac{1}{\sqrt{2}}h_1h_1$,
$h_1h_2$, $\frac{1}{\sqrt{2}}h_2h_2$.

The most stringent unitarity bound on the $g'_1$ coupling is derived
from the requirement that the magnitude of the largest eigenvalue
combined with the function $f(g'_1,s;Y^{B-L}_{\chi},x)$ does not exceed
$1/2$.

If we diagonalise the matrix in equation (\ref{a0_matrix}), we find that the
maximum eigenvalue and the corresponding eigenvector are:
\begin{eqnarray}\label{eigenvalue}
\frac{1}{2} \Rightarrow  \frac{1}{2}
\left( z'z' + h_1h_1\sin^2{\alpha}
- h_1h_2\sin{(2\alpha) + h_2h_2\cos^2{\alpha}}
\right).
\end{eqnarray}

Combining the informations of equations (\ref{function})-(\ref{eigenvalue}),
together with the perturbative unitarity condition in
equation (\ref{condition}), we obtain:
\begin{eqnarray}\label{unitarity_condition}
|\textrm{Re}(a_0)|=\frac{(Y^{B-L}_{\chi}g'_1)^2}{32\pi}
\left|
1+2\log{\left( \frac{(Y^{B-L}_{\chi} g'_1x)^2}{s} \right)}
\right|\leq \frac{1}{2}.
\end{eqnarray}

In the last equation, $s$ represents the scale of energy squared at
which the scattering is consistent with perturbative unitarity,
i.e. $s=\Lambda^2$, where $\Lambda$ is
the evolution energy scale cut-off.

Finally, if we consider the contour of this inequality, we find
exactly the boundary condition that solve the set of differential equations in
(\ref{RGE_g2}), giving us the upper limit for $g'_1$ at the EW scale.
In the next section we will combine all these elements to present a 
numerical analysis of the allowed domain of $g'_1$.

%% file: sect_4.tex
The set of differential eqs. (\ref{RGE_g2})
has been evaluated with the well-known Runge-Kutta algorithms and the
unitarity condition has been imposed as a two-point boundary value
with a simple shooting method, that consisted in varying the initial
conditions in dichotomous-converging steps until the unitarity bound
was fulfilled.

\begin{figure}[!ht]
  \begin{center}
        \includegraphics[angle=0,width=0.98\textwidth]{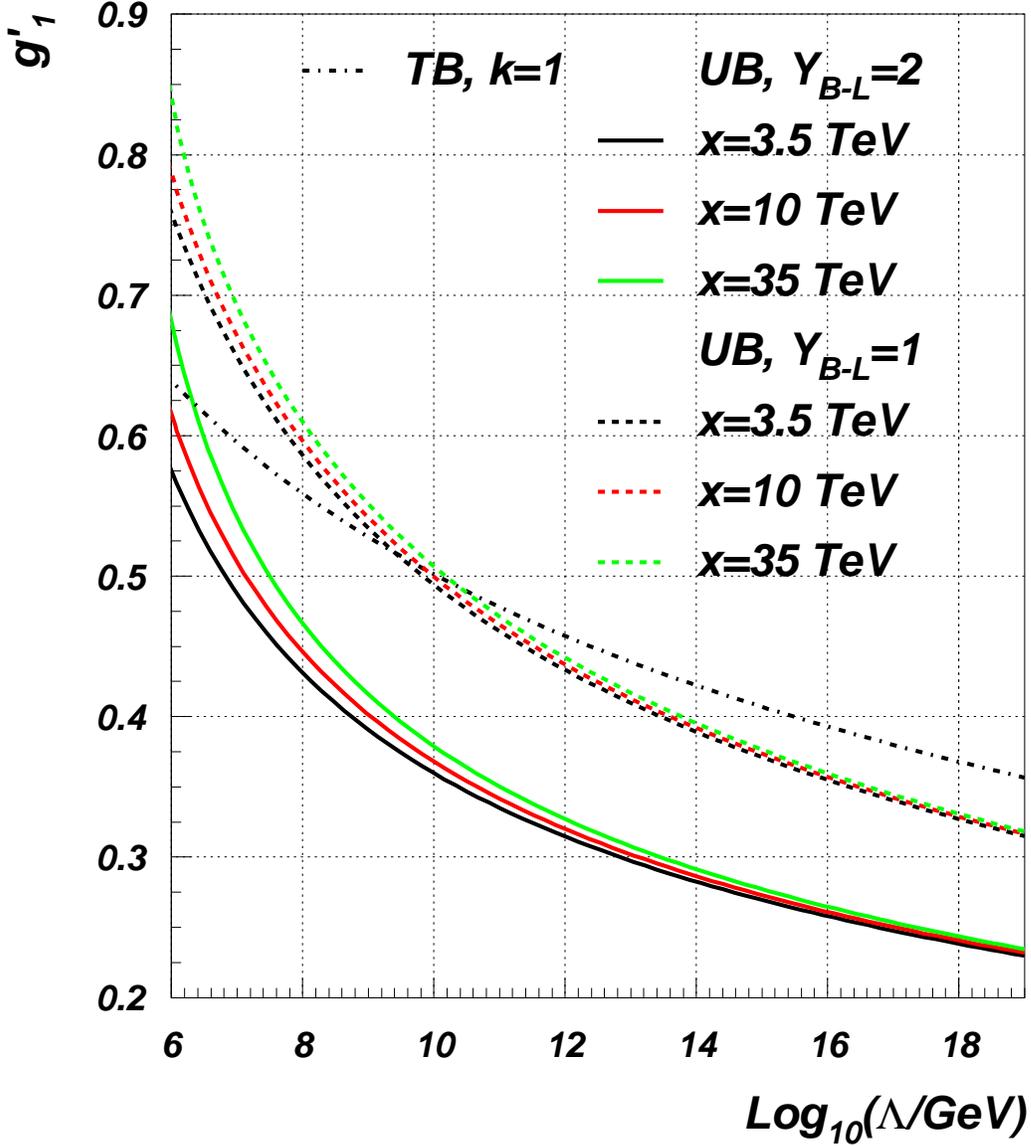}
  \end{center}
  \vspace*{-0.5cm}
  \caption{Triviality (assuming $k=1$, dotted-dashed line) and
        unitarity (continuous and dashed lines) limits on the
        $g'_1$ coupling of
        the minimal $B-L$ model, plotted against the energy cut-off
        $\Lambda$ in $\log_{10}$-scale, for several choices of the
        singlet VEV ($x=3.5$ TeV, black lines; $x=10$ TeV, red (dark
        grey) lines; $x=35$ TeV,
        green (light grey) lines) and the $B-L$ charge of the $\chi$
        field ($Y^{B-L}_{\chi}=2$, continuous-line; $Y^{B-L}_{\chi}=1$,
        dashed-line).}
\label{g1pvsL}
\end{figure}

Moreover, in order to make a fruitful comparison with the ordinary
triviality arguments, we have evaluated the evolution of $g'_1$ with
the two boundary conditions,
equations (\ref{triviality_condition})-(\ref{unitarity_condition}), for
several values of $x$, the Higgs singlet VEV, and
two choices of the $B-L$ charge of the $\chi$ field, corresponding to
the basic model ($Y^{B-L}_{\chi}=2$) and the so-called ``inverse see-saw'' version
($Y^{B-L}_{\chi}=1$) proposed in \cite{Khalil:2010zz}:
the results are plotted in Figure (\ref{g1pvsL}).

In the first place, we verified that the choice of $Y^{B-L}_{\chi}$
(i.e. the choice of model) does not significantly affect the
triviality bounds, so we display them for the case $Y^{B-L}_{\chi}=2$
only.

By direct comparison of the two formulae, it is easy to see that the
unitarity bounds become more important than the triviality bounds when
\begin{eqnarray}
\frac{\Lambda}{x}\simeq
{\rm exp}\left(
\frac{16\pi+(kY^{B-L}_\chi)^2}{(2kY^{B-L}_\chi)^2}
\right),
\end{eqnarray}
with the assumption that $M_{Z'} \sim x$.

From this equation, it is straightforward to see that the choice of
both the $B-L$ charge of the $\chi$ field and the ``ad hoc''
triviality parameter $k$ is crucial for establishing which
limit is the most stringent one.

In the basic version of the model, the choice $Y^{B-L}_\chi=2$ is
necessary to preserve the gauge invariance and we also embrace the
widely accepted assumption $k=1$ as triviality condition. 
If we then choose a value of the VEV $x$ that is compatible with the
experimental limits and 
still in the TeV range, $x\in [3.5,35]$ TeV according
to \cite{Cacciapaglia:2006pk}, we find that the unitarity bounds are
more stringent than the triviality ones when the energy scale is
greater than a 
critical value of $\Lambda_c\simeq 10^6$ GeV, and this is consistent
with the results in Figure (\ref{g1pvsL}).
In a different version of the $B-L$ model, for example 
the ``inverse see-saw'' one \cite{Khalil:2010zz}, where
$Y^{B-L}_\chi=1$, we find that $\Lambda_c=10^9-10^{10}$ GeV, and this
is again confirmed by the plot 
in Figure (\ref{g1pvsL}).

In order to summarise these results, in
Table (\ref{g1p-up_bound}) we present
a comparison between the triviality and the unitarity bounds for several
energy scales and $B-L$-breaking VEVs $x$.

\begin{table}[!ht]
\begin{center}
\begin{tabular}{|c|c|c|c|c|c|}
\hline
$Log_{10} (\Lambda/ \mbox{GeV}) $ & 7 & 10 & 15 & 19  \\
\hline
TB, $g_1'(k=1)$  & 0.595 & 0.501 & 0.407 & 0.357  \\ 
\hline
UB, $g_1'(x=3.5 \ \mbox{TeV})$  & 0.487 & 0.360 & 0.269
& 0.230  \\ 
\hline
UB, $g_1'(x=10 \ \mbox{TeV})$  & 0.510 & 0.368 & 0.273
& 0.232  \\ 
\hline
UB, $g_1'(x=35 \ \mbox{TeV})$  & 0.542 & 0.379 & 0.277
& 0.234  \\ 
\hline
\end{tabular}
\end{center}
\vskip -0.5cm
\caption{Triviality bounds,
equation (\ref{triviality_condition}) with $k=1$, and unitarity bounds,
equation (\ref{unitarity_condition}) with $x=3.5,10,35$ TeV, for $g'_1$
in the standard $B-L$ model for several values of the
energy scale $\Lambda$.
\label{g1p-up_bound}}
\end{table}

Though these results are scale dependent, we see that, if $\Lambda
\gg \Lambda_c$, our method basically refines the triviality bound by an
absolute value of $\simeq 0.1$, that represents a correction of (at
least) $20\%$ on the results that have recently appeared in the literature
(see \cite{Basso:2010jm}).

%% file: sect_5.tex
In this paper, we have shown that, by combining perturbative unitarity and RGE
methods, one can significantly constrain the
$g'_1$ coupling of the minimal $B-L$ extension of the SM, by imposing
limits on its upper value
that are more stringent than standard triviality bounds.
(Also notice that, as unitarity is more constraining than triviality,
the stability of the perturbative solution obtained through the former is
already guaranteed by the latter.)

We presented a full set of analytical results, plus a significative comparison between the type-I see-saw and the "inverse" see-saw (neutrino mass generation mechanisms) implementation, that turned out to be analytically accessible in the minimal $B-L$ (i.e., no mixing between $Z_{SM}$ and $Z_{B-L}$ at tree-level) extension of the $SM$.

Finally, we have verified by direct computation that even if a reasonably small $\widetilde{g}$ (e.g., $|\tilde{g}<0.05|$) is switched on, our conclusions are unchanged.

The present work, alongside Refs.~\cite{Basso:2010jt}
and \cite{Basso:2010jm}, enables one to ultimately define the combined experimental and theoretical
limits on the Higgs sector of the minimal $B-L$ model, in view of its exploration at present and future colliders
\cite{Higgs_At_Colliders}.

%% file: acknowledgements.tex
GMP would like to thank Alexander S.~Belyaev, Stephen F.~King and
Douglas A. Ross for helpful discussions.